\documentclass[symmetry,article,accept,oneauthor,pdftex]{Definitions/mdpi}

\newcommand{\aap}{{\it Astron. Astrophys.}}

\newcommand{\araa}{{\it Annu. Rev. Astron. Astrophys.}}
\newcommand{\apj}{{\it Astrophys. J.}}
\newcommand{\apjl}{{\it Astrophys. J. Lett.}}
\newcommand{\apjs}{{\it Astrophys. J. Supp. Series}}
\newcommand{\apss}{{\it Astrophys. Space Sci.}}
\newcommand{\jgr}{{\it J. Geophys. Res.}}

\newcommand{\mnras}{{\it Mon. Not. Roy. Astron. Soc.}}
\newcommand{\nat}{{\it Nature}}

\newcommand{\prl}{{\it Phys. Rev. Lett.}}
\newcommand{\solphys}{{\it Solar Phys.}}

\newcommand{\ssr}{{\it Space Sci. Rev.}}

\firstpage{1} 
\pubvolume{1}
\issuenum{1}
\articlenumber{0}
\pubyear{2025}
\copyrightyear{2025}
\datereceived{11 December 2024} 
\daterevised{21 January 2025 }
\dateaccepted{22 January 2025} 
\datepublished{} 
\hreflink{https://doi.org/} 

\usepackage{todonotes}
\usepackage{caption,subfigure}
\usepackage{bm}
\usepackage{subfigure}
\makeatletter
\renewcommand{\@thesubfigure}{\normalsize(\textbf{\alph{subfigure}})}
\makeatother
\Title{Logarithmic Separable Solutions of Force-Free Magnetic Fields in Plane-Parallel and Axial Symmetry}

\TitleCitation{Logarithmic Separable Solutions of Force-Free Magnetic Fields in Plane-Parallel and Axial Symmetry}

\Author{Konstantinos N. Gourgouliatos
 \orcidA}

\AuthorNames{Konstantinos N. Gourgouliatos}

\AuthorCitation{Gourgouliatos,
 K.N.}

\address[1]{%
Department
 of Physics, University of Patras, 26504 Patras, Greece; kngourg@upatras.gr}

\abstract{This work introduces a systematic method for identifying analytical and semi-analytical solutions of force-free magnetic fields with plane-parallel and axial symmetry. The method of separation of variables is used, allowing the transformation of the non-linear partial differential equation, corresponding to force-free magnetic fields, to a system of decoupled ordinary differential equations, which nevertheless, are in general non-linear. It is then shown that such solutions are feasible for configurations where the electric current has a logarithmic dependence to the magnetic field flux. The properties of the magnetic fields are studied for a variety of physical parameters, through solution of the systems of the ordinary differential equations for various values of the parameters. It is demonstrated that this new logarithmic family of solutions has properties that are highly distinct from the known linear and non-linear equations, as it allows for bounded solutions of magnetic fields, for periodic solutions and for solutions that extend to infinity. Possible applications to astrophysical fields and plasmas are discussed as well as their use in numerical studies, and the overall enrichment of our understanding of force-free configurations. }


\begin{document}


\section{Introduction}

Magnetic fields mark their presence across the universe being key elements in stars~\cite{2009ARA&A..47..333D}, the~interstellar medium~\cite{2010ApJ...725..466C,2021A&A...647A.186S} and the intergalactic medium~\cite{2017ARA&A..55..111H}, but~also in space~\cite{1990JGR....9511957L} and the vicinity of the Sun~\cite{1961ApJ...133..572B,2016SSRv..204...49B}. Magnetic fields are even more critical, providing in several cases, an~avenue of observation for compact objects, and~they have been observed in white dwarfs, neutron stars and in the vicinity of black holes~\cite{1981ApJS...45..457A,2005MNRAS.356.1576W,2018ASSL..457...57G,2013Natur.501..391E}. While in most cases magnetic fields are dynamically subdominant and provide only modifications to the overall equilibrium; for instance, the internal structure and equilibrium of stars is determined by gravity and pressure and magnetic fields may lead to small deviations from spherical shape~\cite{2001ApJ...554..322C,2014MNRAS.437..424L,2022A&A...666A.138L},  
 there are several astrophysical systems where magnetic fields have a central role in driving the evolution of the system and dictate the overall equilibrium. Such systems include the solar corona, where the field has been reconstructed considering a dynamically dominant magnetic field~\cite{2006SoPh..235..161S,2008SoPh..247..269M,1997SoPh..174..129A}, magnetospheres of pulsars~\cite{1973ApJ...182..951S} and, in~general,  regions where the thermal pressure is negligible compared to the magnetic pressure. Beyond~astrophysics, strong magnetic field configurations are relevant to plasma confinement in fusion devices~\cite{seki1985stability,1996ffmg.book.....M} and laboratory plasmas where strong magnetic fields which govern the overall dynamics of the system~\cite{1990PhRvL..65.2011S}.  Thus, the~study of steady-states dominated by strong magnetic fields is critical for modeling and eventual understanding of systems, ranging from astrophysical scales to laboratory applications.  

 Considering a highly conducting plasma where the magnetic field dominates, one can prescribe its evolution and equilibrium based on its magnetic field. Owing to high conductivity, the~plasma is frozen into the magnetic field. The~dominant force in the system is the magnetic one, thus while it acts onto the magnetic field it drags the plasma along, whose inertia is small. If~the system reaches an equilibrium it will be such that the magnetic force vanishes. Apart from the trivial, current-free solution, where the magnetic field is potential and satisfies Laplace's equation, a~far more interesting case arises when the field arranges itself in a way that the current is parallel to the magnetic field. Such an equilibrium is called~force-free. 

Despite its conceptual simplicity, the~specification of force-free solutions is complicated, as~the function providing the electric current is not determined in advance. Analytical solutions have been found for a limited number of configurations. A~class of systems that has been thoroughly explored is the linear force-free magnetic fields in non-relativistic systems~\cite{1956PNAS...42....1C,1957ApJ...126..457C,1956ApJ...124..232C,1951PhRv...83..307L,1989AuJPh..42..317D,1999ApJ...518..948W,2019A&C....26...50B}, but~also in relativistic regimes~\cite{2005MNRAS.359..725P,2008MNRAS.391..268G,2009MNRAS.396.2399G,2022ApJ...934..140B}.  In linear force-free fields, in~addition to the necessary condition that the current is parallel to the magnetic field, it is assumed that the ratio of the modulus of the electric current to the magnetic field is constant throughout the entire domain. This simplifies the partial differential equation to a linear form that can be solved analytically subject to given boundary conditions. This however comes at the cost, introducing a particular length-scale, which is directly related to the aforementioned ratio. Under~this assumption, the~field creates disconnected regions where the mathematical equations are satisfied, but~the field is no longer in contact with { the astrophysical object is originates from}, i.e.,~the star where it emanates from~\cite{2024MNRAS.527.6691N}, or~the source of the outflow~\cite{1990ApJ...348L..73Y,2010GApFD.104..431G}. Classes of non-linear solutions have been found in the form of self-similar force-free configurations~\cite{1994A&A...288.1012A,1994MNRAS.267..146L,1996Ap&SS.243...29B,2008MNRAS.385..875G,2008MNRAS.391..268G}, where the electric current function obeys a power-law relation with the magnetic flux. This family of solutions allows for configurations that reach infinity and do not have a particular scale where they become zero, which allows for broader applications to systems where the magnetic fields are expected formally to extend to infinity. This comes at the expense of a particular functional relation, which in general is non-linear. Such non-linear solutions have been used in the modeling of solar magnetic fields~\cite{2008JGRA..113.3S02W,2008SoPh..247..269M,2018SSRv..214...99Y}. 

In the present work, a~systematic approach for separable force-free configurations is presented, following the basic steps of an approach discussed in~\cite{2009PhDT.......295G}. Apart from recovering the known linear solutions, a~family of previously unknown solutions is found, the~logarithmic ones, which are studied in~detail.  

The plan of the paper is the following. In~Section~\ref{Sec_FF}, the~basic properties of force-free magnetic fields and the corresponding mathematical setup, are reviewed. In~Section~\ref{sec:PP}, separable solutions for plane-parallel force-free magnetic fields are presented in Cartesian geometry. In~Section~\ref{sec:cyl}, I present solutions for force-free magnetic fields in cylindrical geometry under azimuthal symmetry. The~overall properties of the logarithmic solutions are discussed in Section~\ref{sec:disc}. I conclude in Section~\ref{sec:Con}.

\section{Force-Free Magnetic~Fields}
\label{Sec_FF}

The ratio of the plasma thermal pressure to the magnetic pressure is quantified through the plasma $\beta$ parameter~\cite{1994ppit.book.....S}: 
\begin{eqnarray}
    \beta = \frac{p}{p_{mag}}=\frac{8 \pi n kT}{B^2}\,,
\end{eqnarray}
where $p$ is the gas thermal pressure and $p_{mag}$ the magnetic pressure, $n$ is the particle number density of the plasma, $k$ is the Boltzmann constant, $T$ is the temperature and $B$ the magnetic field. Thus, for~systems with $\beta\ll 1$ the magnetic field is the prime driver of the evolution as it dominates over the thermal pressure. Furthermore, the~plasma is frozen to the fluid due to its high conductivity~\cite{2005ppa..book.....K}. Should the system be able to reach a { steady-state}, this will be determined by the magnetic field and it will  be such that the magnetic field is parallel to the { electric} current so that

\begin{eqnarray}
    {\bf J} \times {\bf B} = {\bf 0}\,,
    \label{eq:FF0}
\end{eqnarray}
where ${\bf J}$ is the electric current density. Note that this holds in the limit of non-relativistic fields. The~electric current is related to the magnetic field through Amp\`ere's law:
\begin{eqnarray}
{\bf J}=\frac{4 \pi}{c} \nabla \times {\bf B}\,.
\label{eq:AMPERE}
\end{eqnarray}

Combining
 Equation~(\ref{eq:AMPERE}) with \eqref{eq:FF0} one obtains the following expression:
\begin{eqnarray}
    \left(\nabla \times {\bf B}\right) \times {\bf B}={\bf 0}\,,
    \label{eq:FF1}
\end{eqnarray}
which is equivalent to the condition that the magnetic field parallel to its curl:
\begin{eqnarray}
    \nabla \times {\bf B}=\alpha {\bf B}\,.
    \label{eq:CurlB-B}
\end{eqnarray}

Taking the divergence of the above expression a relation between $\alpha$ and the magnetic field ${\bf B}$ is found:
\begin{eqnarray}
{\bf B} \cdot \nabla \alpha = {\bf 0}\,,
\label{eq:alpha}
\end{eqnarray}
by virtue of Gauss' law for magnetism $\nabla \cdot {\bf B} = 0$ and the vector calculus identity that the divergence of the curl is zero. Equation~(\ref{eq:alpha}) dictates that $\alpha$ must be a constant along a magnetic field line as it states that the directional derivative of $\alpha$ along ${\bf B}$ is~zero. 

\section{Plane Parallel Force-Free Magnetic~Fields}

\label{sec:PP}

Consider a magnetic field in Cartesian geometry $(x,y,z)$ which is symmetric under translations parallel to the $z$ axis. This magnetic field can be written in terms of two scalar functions $\Psi(x,y)$ and $B_z(x,y)$ as follows:
\begin{eqnarray}
    {\bf B} = \nabla \Psi \times \nabla z + B_{z}\nabla z\,.
\end{eqnarray}

The above expression satisfies the zero divergence condition by construction. Substituting the magnetic field to the force-free Equation~(\ref{eq:FF1}), the~following equation from the $z$ component is found:
\begin{eqnarray}
    \frac{\partial \Psi}{\partial x} \frac{\partial B_z}{\partial y}-\frac{\partial B_z}{\partial x} \frac{\partial \Psi}{\partial y} =0\,,
\end{eqnarray}
\textls[-10]{which is the Jacobian of $\Psi$ and $B_z$ with respect to $x$ and $y$. As~it is equal to zero, $B_z=B_{z}(\Psi)$. Then, from~the $x$ and $y$ components of the force equation one obtains the following~expression:}
\begin{eqnarray}
    \frac{\partial^2 \Psi}{\partial x^2} +\frac{\partial^2 \Psi}{\partial y^2} = -B_z B_z^{\prime}\,.
    \label{eq:PDE}
\end{eqnarray}

Hereafter, a~prime denotes differentiation with respect to the quantity a function depends on, in~the particular case it is $\Psi$, as~it has been found that $B_z$ is a function of $\Psi$. Next, a~separable solution is assumed, so that $\Psi (x,y) = X(x) Y(y)$. Substituting the separable solution and on division with $\Psi$, Equation~(\ref{eq:PDE}) takes the following form: 
\begin{eqnarray}
    \frac{X^{\prime \prime}}{X} + \frac{Y^{\prime \prime}}{Y} = -\frac{B_z B_z^{\prime}}{\Psi}\,. 
    \label{eq:sep}
\end{eqnarray}

One can define the following functions:
\begin{eqnarray}
    F(X) = \frac{X^{\prime \prime}}{X}\,,\nonumber \\
    G(Y) = \frac{Y^{\prime \prime}}{Y}\,,\nonumber \\
    H(\Psi) = \frac{B_z B_z^{\prime}}{\Psi}\,. \nonumber 
\end{eqnarray}

Then, by~substituting into Equation~(\ref{eq:sep}), the~following equation is found:
\begin{eqnarray}
    F(X) +G (Y) =-H(\Psi)\,,
    \label{eq:FGH}
\end{eqnarray}
and acting with the operator $X \frac{d}{dX}$, the~equation takes the following form:
\begin{eqnarray}
    X \frac{dF}{dX} &=& -X \frac{dH}{d\Psi}\frac{d\Psi}{dX} \nonumber \\
    X\frac{dF}{dX} &=& -\Psi \frac{dH}{d\Psi} \nonumber \\
    \frac{dF }{d \ln X} &=& -\frac{dH}{d \ln \Psi}=c_0\,.
\end{eqnarray}

The last expression is equal to an arbitrary constant $c_0$, as~the first part is a function of $X$ only, and~the middle only a function of $\Psi$. Note that, if~$Y(y)$ is a constant there is no requirement to set the above expression equal to $c_0$, as~$\Psi=\Psi(X)$, but~such a field will depend only on $x$ and correspond to a simpler one-dimensional problem, whereas in the current work the focus is on fields that depend on both~coordinates. 

Integrating the equation for $H(\Psi)$, one obtains:
\begin{eqnarray}
    H = -c_0 \ln \Psi +c_1 \,
    \label{eq:H}
\end{eqnarray}
then substituting back for $B_z$:
\begin{eqnarray}
    \frac{1}{2} \left(B_z^2\right)^{\prime}=- c_0\Psi \ln \Psi + c_1 \Psi \,,
\end{eqnarray}
which can be directly integrated, and~eventually takes the form:
\begin{eqnarray}
    B_z^2 = \left(-c_0 \ln \Psi +\frac{c_0}{2}+c_1\right)\Psi^2 +c_2\,.
    \label{eq:Bz}
\end{eqnarray}

This is the general form of $B_z$ that allows separable solutions. In~this expression, three free constants appear: $c_0$, $c_1$ and $c_2$. They are not predetermined, but~they need to have values so that the right-hand-side of the equation is a non-negative number as it is equal to the square of a real number. {  Note that the derivation of the force-free partial differential equation can also start from the expression \eqref{eq:alpha}, where the value of $\alpha$ is also determined in terms of the flux function. This proccess is shown in Appendix \ref{appA}.} 

To proceed with the solution, one can substitute the expression of Equation~(\ref{eq:H}) into Equation~(\ref{eq:FGH}) and subsequently obtain:
\begin{eqnarray}
    \frac{X^{\prime \prime}}{X} +\frac{Y^{\prime \prime}}{Y} =c_0 \ln X +c_0 \ln Y -c_1\,,
\end{eqnarray}
which can be rearranged into the following form and set equal to a constant $k_0$, as~both the left part and the middle part of the equation depend only on $X$ and $Y$, respectively:
\begin{eqnarray}
    \frac{X^{\prime \prime}}{X}-c_0 \ln X = -\frac{Y^{\prime \prime}}{Y}+c_0 \ln Y -c_1=k_0\,.
\end{eqnarray}

These equations can be integrated analytically once, leading to the following\linebreak expressions:
\begin{eqnarray}
    X^{\prime 2}-c_0X^2\left(\ln X -\frac{1}{2}\right)-k_0X^2+d_1=0\,,\nonumber \\
    Y^{\prime 2}-c_0Y^2\left(\ln Y -\frac{1}{2}\right)+\left(k_0+c_1\right)Y^2+d_2=0\,,
    \label{eq:XY-prime}
\end{eqnarray}
where $d_1$ and $d_2$ are integration constants. The~above expressions do not have further obvious analytical solutions, yet their numerical integration is straightforward for appropriate boundary~conditions.

One can notice that for $c_0=0$, the linear force-free solution is recovered admitting solutions in the form of trigonometrical or hyperbolic functions, depending on the choice of the sign of $c_1$ and $k_0$.  

To unveil the qualitative characteristics of the logarithmic solution, I set $k_0=0$ so that the linear term vanishes. Given that $X$ and $Y$ obey the same equation subject to a difference of the constant $c_1$, the~focus will be on $X(x)$ and the main conclusions can also be applied to $Y(y)$. Throughout this work the ordinary differential equations appearing have been integrated using the Runge--Kutta 4th order method and the solutions have been tested for convergence by using half step size to ensure the changes are beyond the sixth significant~digit. 

\begin{figure}[H]
    
    \includegraphics[width=0.99\linewidth]{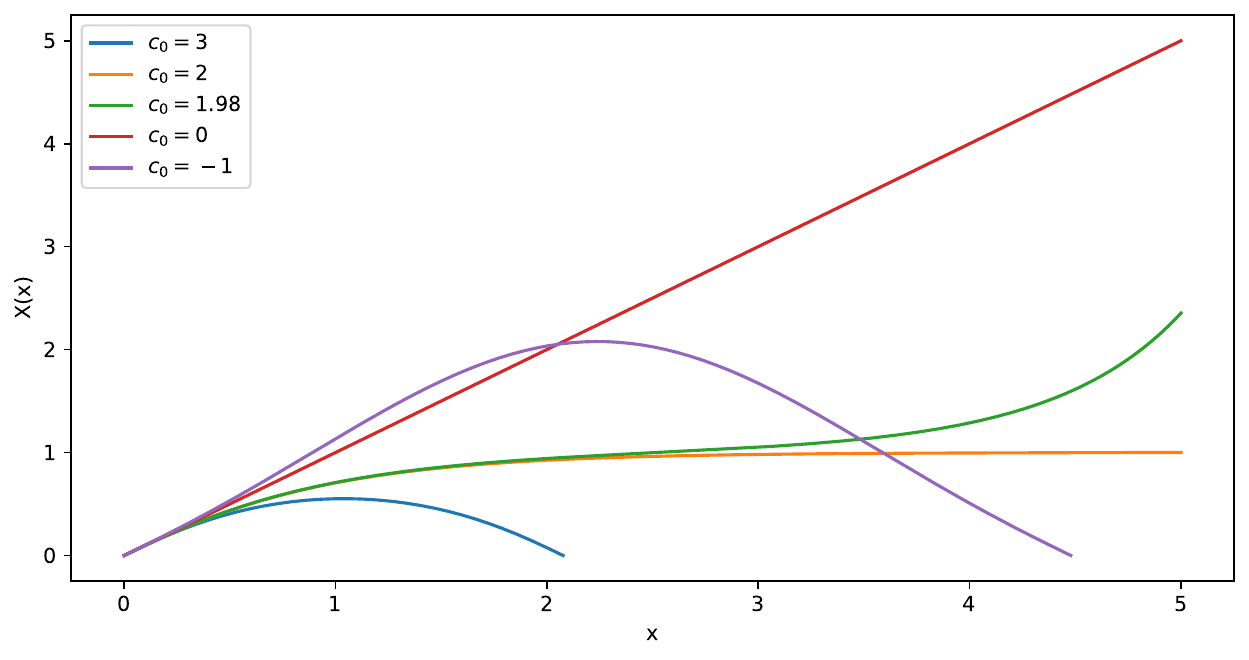}
    \caption{The solution for $X(x)$ for the following set of parameters: $c_0=-1,~0,~1.98,~2,~3$, while in all solutions $k_0=0$, $c_1=0$, $d_1=1$, and~boundary condition $X(0)=0$. The~curves demonstrate the qualitative difference of the solution depending on the value of $c_0$ multiplying the logarithmic~term. }
    \label{fig:X_Logarithmic}
\end{figure}

There is the requirement that $X\geq 0$ so that the $X\ln X$ is real-valued. First, it is assumed that $X(0)=0$, thus the first derivative $X^{\prime}$ needs to be positive, so it is chosen that $X^{\prime}(0)=1$, which also determines the value of the constant $d_1=1$. Near~the origin $X^{\prime}>0$, while $X^{\prime \prime}<0$. To~make both the first and the second derivatives equal to zero, it has to be $X(x_c)=1$, $c_0=2$. If~this is the case, for~$x>x_c$ the function adopts the constant value of unity, thus $X(x>x_c)=1$, Figure~\ref{fig:X_Logarithmic}, orange curve. Whereas for $c_0>2$, as~$X$ never exceeds unity, thus the term $X\ln X$ is negative throughout the integration, leading to the decrease of $X$ until it becomes zero,   Figure~\ref{fig:X_Logarithmic}, blue curve. For~$0<c_0<2$, at~some point $X$ reaches unity, while its derivative is still positive, so $X\ln X$ becomes positive and leads to rapid growth of $X$ beyond that point, Figure~\ref{fig:X_Logarithmic}, green curve. For~$c_0=0$ the derivative remains constant and the system admits a linearly increasing solution, Figure~\ref{fig:X_Logarithmic}, red curve. Finally, for~$c_0<0$, as~the term $c_0 X \ln X$ is always negative, $X(x)$ reaches a maximum, then decreases and becomes zero at some point depending on the value of $c_0$, Figure~\ref{fig:X_Logarithmic}, purple~curve. 

\begin{figure}[H]
    
    \includegraphics[width=0.99\linewidth]{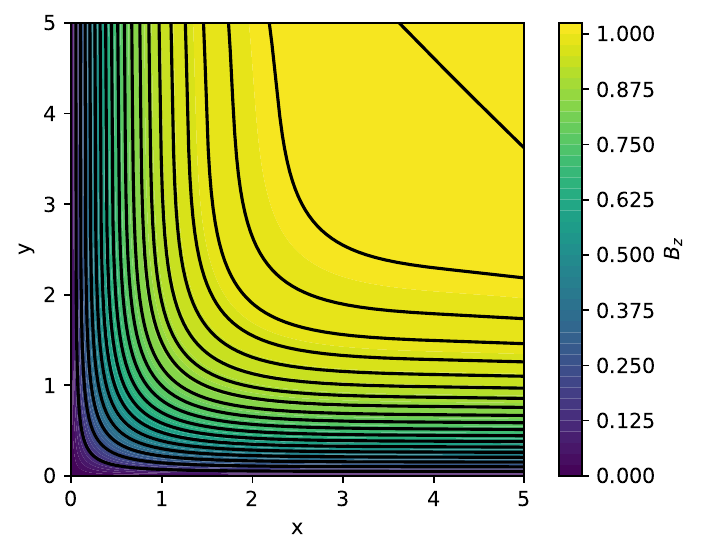}
    \caption{The structure of the magnetic field corresponding to the logarithmic magnetic field solution, for~the solution corresponding to $c_0=2$, $k_0=c_1=c_2=0$. The~field structure due to the $B_x$ and $B_y$ components of the field is  
shown in the form of black contours of constant $\Psi$. The~density of the black contours corresponds to the intensity of the $B_x$ and $B_y$ field. The~$B_z$ component is shown in color. The~$B_x$ and $B_y$ components field for large $x$ and $y$ tend to zero, while $B_z$ adopts a constant~value. }
    \label{fig:Field_Lines_Plane-Flat}
\end{figure}

The field line structure of the system where $\Psi$ becomes constant and equal to unity for sufficiently large $x$ and $y$, having $c_0=2$, corresponding to the solution of the orange  curve
of Figure~\ref{fig:X_Logarithmic} is shown in Figure~\ref{fig:Field_Lines_Plane-Flat}, with~the $B_z$ component shown in Figure~\ref{fig:Bz-Flat}.

\vspace{-12pt}

\begin{figure}[H]
    
    \includegraphics[width=0.99\linewidth]{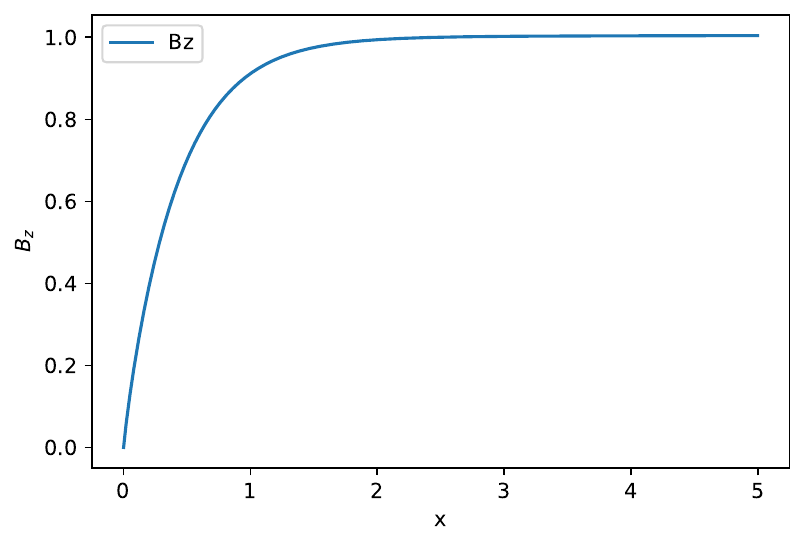}
    \caption{The $B_{z}$ component of the magnetic field at $y=1.7$ using the same parameters as in the logarithmic solution shown in Figures~\ref{fig:Field_Lines_Plane-Flat}. It is evident that the $B_{z}$ becomes constant at large $x$.}
    \label{fig:Bz-Flat}
\end{figure}

Upon consideration of the linear term $k_0\neq 0$, the~solution is modified.
An example of solutions is shown in Figure~\ref{fig:XY_solution}, where the logarithmic solution is shown in blue for $c_0=-1$ and the linear solution for $c_0=0$ is shown in orange. The~other parameters for both solutions are $k_0=-1$, $c_1=2$, $d_1=1$, $d_2=1$ and boundary condition $X(0)=0$, $Y(0)=0$. It is noticeable that once the logarithmic term is present, due to its impact it affects the maxima and the roots of the solution. Furthermore, due to the presence of the logarithm, negative values of $\Psi$ are not acceptable, thus once the solution becomes zero it cannot continue any further, which is the case for the logarithmic solution at approximately $x=3.2$. The~structure of the field at the $x-y$ plane is shown in Figure~\ref{fig:Field_Lines_Plane}, while the $B_{z}$ component for this solution as a function of $x$, for~a section at $y=1.6$ is shown in Figure~\ref{fig:Bz}.

\begin{figure}[H]
    
    \includegraphics[width=0.99\linewidth]{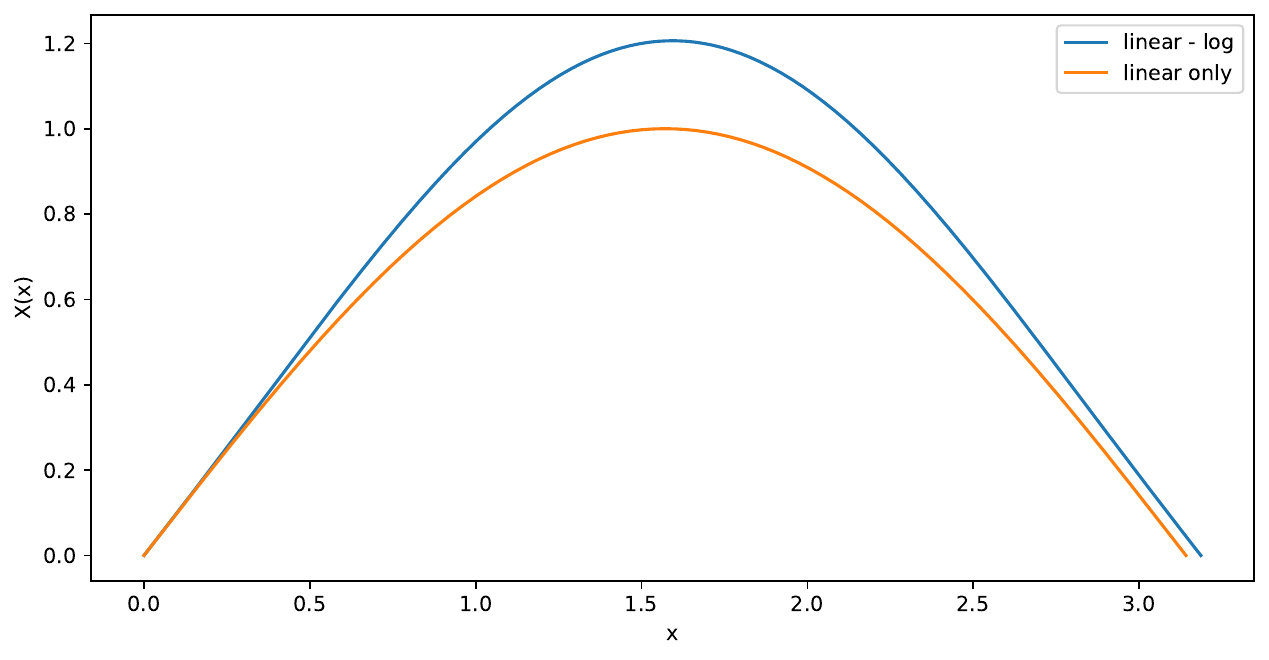}
\caption{\textit{Cont.}}
\end{figure}
\begin{figure}[H]\ContinuedFloat
    \includegraphics[width=0.99\linewidth]{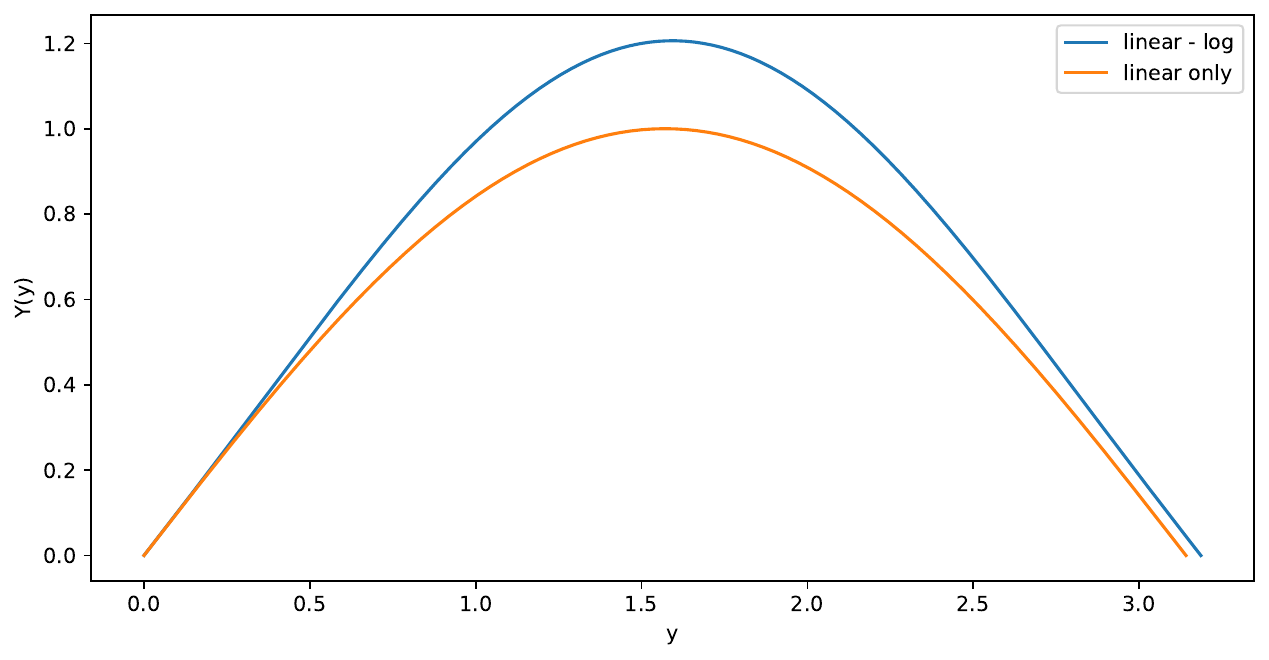}
    \caption{The solution for the $X(x)$ (top) and $Y(y)$ (bottom) part of the solution following the integration of Equation~(\ref{eq:XY-prime}). The~blue curves corresponds to the logarithmic solution for the following choice of parameters: $c_0=-1$, $k_0=-1$, $c_1=2$, $d_1=1$, $d_2=1$ and boundary condition $X(0)=0$, $Y(0)=0$. The~orange curves correspond to the linear solution where $c_0=0$ and all other parameters are the same as in the logarithmic~solution.}
    \label{fig:XY_solution}
\end{figure}

\vspace{-15pt}

\begin{figure}[H]
    
    \includegraphics[width=0.99\linewidth]{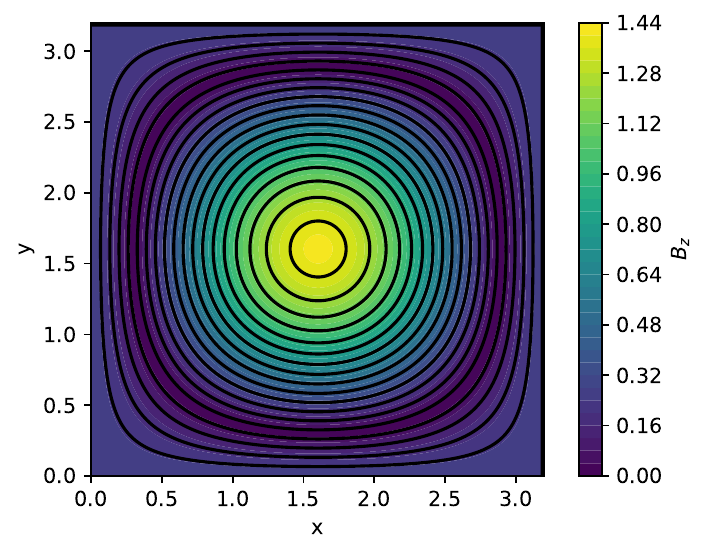}
    \caption{The structure of the magnetic field corresponding to the logarithmic magnetic field solution, for~the solution and choice of parameters shown in Figure~\ref{fig:XY_solution} while the additive constant $c_2=0.068$ to avoid negative values in the square root. The~field structure due to the $B_x$ and $B_y$ components of the field  is
 shown in the form of black contours of constant $\Psi$. The~density of the black contours corresponds to the intensity of the $B_x$ and $B_y$ field. The~$B_z$ component is shown in~color.}
    \label{fig:Field_Lines_Plane}
\end{figure}
\unskip
\begin{figure}[H]
    
    \includegraphics[width=0.99\linewidth]{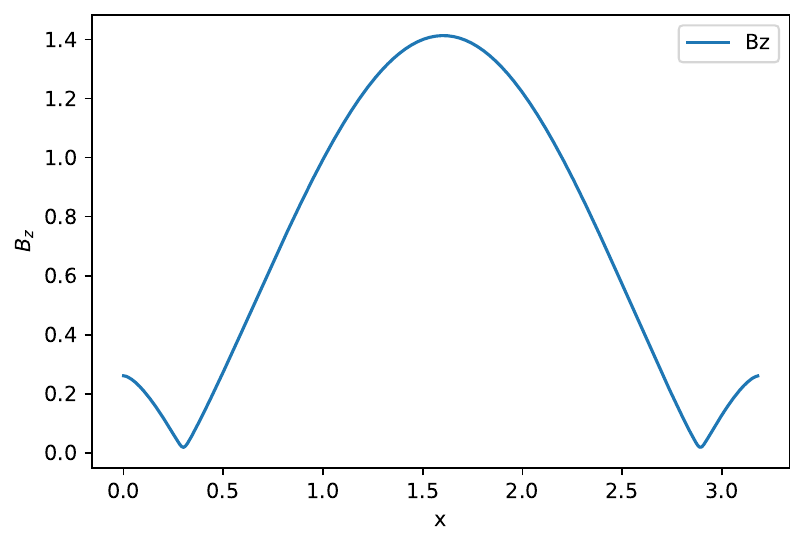}
    \caption{The $B_{z}$ component of the magnetic field at $y=1.7$ using the same parameters as in the logarithmic solution shown in Figures~\ref{fig:XY_solution} and \ref{fig:Field_Lines_Plane}. }
    \label{fig:Bz}
\end{figure}

\begin{figure}[H]
    
    \includegraphics[width=0.99\linewidth]{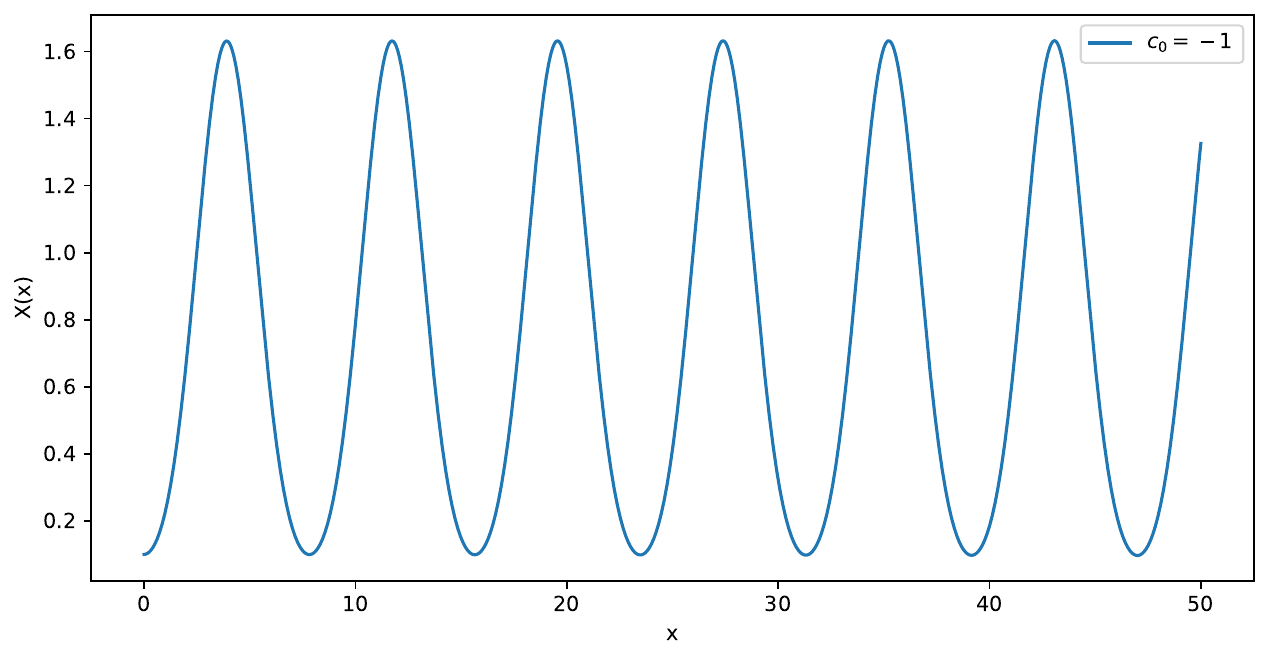}
    \caption{The periodic solution for $X(x)$ that has a periodic behavior, for~$c_0=-1$, with~boundary conditions $X(0)=0.1$ and $X^{\prime}(0)=0$, while $k_0=c_1=0$ and $d_1=0$ so that it is compatible with the zero derivative boundary~condition.  }
    \label{fig:X-oscillatory}
\end{figure}

\begin{figure}[H]
    
    \includegraphics[width=0.99\linewidth]{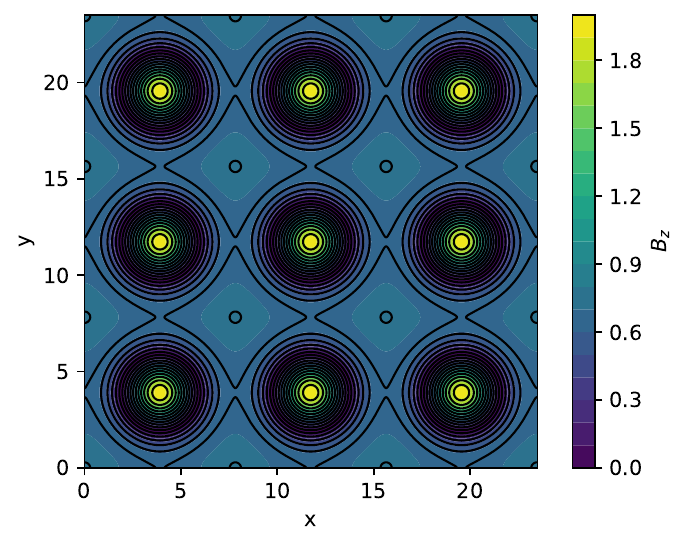}
    \caption{The field line structure, with~the $B_x - B_y$ field lines shown in black and $B_z$ in color for the periodic solution corresponding to the parameters of Figure~\ref{fig:X-oscillatory}. }
    \label{fig:Plane-oscillatory}
\end{figure}

A qualitatively different family of solutions demonstrates periodic behavior. Setting the boundary condition $X(0)=0.1$, $X^{\prime}=0$, and~the constant $c_0=-1$, the~solution oscillates between a maximum and a minimum value, both of them being positive as shown in Figures~\ref{fig:X-oscillatory} and \ref{fig:Plane-oscillatory}.

\unskip

\unskip

\section{Cylindrical Fields with Azimuthal~Symmetry}
\label{sec:cyl}

Next, I will consider systems in cylindrical geometry $(R, \phi, z)$ under axial symmetry, so that there is no dependence on $\phi$. Such magnetic fields can be expressed in terms of two scalar functions:
\begin{eqnarray}
    {\bf B} = \nabla P(R,z) \times \nabla \phi + T(R,z) \nabla \phi\,.
\end{eqnarray}

Applying the force-free condition, Equation~(\ref{eq:FF1}), one obtains the following two\linebreak expressions:
\begin{eqnarray}
    \frac{\partial P}{\partial z}\frac{\partial T}{\partial R} - \frac{\partial T}{\partial z}\frac{\partial P}{\partial R}=0\,,\nonumber \\
    R\frac{\partial }{\partial R} \left(\frac{1}{R}\frac{\partial P}{\partial R}\right)+\frac{\partial^2 P}{\partial z^2} = -TT^{\prime}\,.
\end{eqnarray}

From the first equation, being the Jacobian of $P$ and $T$ with respect to $R,z$ it is evident that $T = T(P)$, which is then used in the second equation, where the right-hand-side contains the derivative of $T$ with respect to $P$. Upon~assumption of a separable solution $P(R,z) = V(R) Z(z)$ and on division with $P$ the expression becomes:
\begin{eqnarray}
    \frac{R}{V}\frac{d}{dR}\left(\frac{1}{R}\frac{dV}{dR}\right)+\frac{Z^{\prime \prime }}{Z}=-\frac{TT^{\prime}}{P}\,.
\end{eqnarray}

In the first term of the left-hand-side of the equation, only the function $V(R)$ appears and its derivatives, in~the second only $Z(z)$, and~the right-hand-side is only a function of $P$. Applying the technique used in the plane-parallel problem, the~equation reduces to the following expression:
\begin{eqnarray}
    K(V) + L(Z) = -M(P)\,,
\end{eqnarray}
where:
\begin{eqnarray}
    K(V) = \frac{R}{V}\frac{d}{dR}\left(\frac{1}{R}\frac{dV}{dR}\right)\,,\nonumber \\
    L(Z) = \frac{Z^{\prime \prime }}{Z} \,, \nonumber \\
    M (P) = \frac{TT^{\prime}}{P}\,.
\end{eqnarray}

Acting now with the operator $Zd/dZ$, the~equation becomes:
\begin{eqnarray}
    \frac{d L}{d \ln Z} = -\frac{dM}{d \ln P} = a_0\,,
\end{eqnarray}
where $a_0$ is a constant. Integrating the second equality of the above equation and substituting the expression for $T$, the~following form for $TT^{\prime}$ is found:
\begin{eqnarray}
    TT^{\prime} =\frac{1}{2} \left(T^2\right)^{\prime}= -a_0 P \ln P + a_1 P\,.
\end{eqnarray}
where $a_1$ is a constant. This equation, upon~integration, provides the following expression for $T(P)$: 
\begin{eqnarray}
    T^2 =\left(-a_0 \ln P +\frac{a_0}{2}+a_1\right)P^2 +a_2\,.
    \label{eq:T-sol}
\end{eqnarray}

Then, the~equation for $V$ and $Z$ reduces to the following form:
\begin{eqnarray}
    \frac{R}{V} \frac{d}{dR}\left(\frac{1}{R} \frac{dV}{dR}\right) -a_0 \ln V =-\frac{Z^{\prime \prime}}{Z} + a_0 \ln Z -a_1 = k_1\,, 
\end{eqnarray}
where $k_1$ is a constant. Separating the equations, they become:
\begin{eqnarray}
    V^{\prime \prime} -\frac{1}{R}V^{\prime}-a_0 V \ln V - k_1 V=0\,, \nonumber \\
    Z^{\prime \prime} -a_0 Z \ln Z +(a_1 +k_1) Z =0\,.
\end{eqnarray}

{As in the plane-parallel case, this derivation can start from Equation~(\ref{eq:alpha}), the~relevant steps are shown in the Appendix 
 \ref{appA}.} 

The decoupled equations can be integrated numerically subject to a set of appropriate boundary conditions. It comes to little surprise that the equation for $Z(z)$ bears no difference from the plane-parallel equations. Focusing on the solution of the radial part of the equation, the~physically accepted boundary conditions are those where $V(0)=0$ and $\lim_{R\to 0} \frac{V(0)^{\prime}}{R} \in \mathbb{R}$. This also affects the solution of the toroidal field: since it needs to remain finite it must be $\lim_{R\to 0} \frac{T}{R}\in \mathbb{R}$, if~this limit is non-zero then a line current flows along the axis, while if it is zero, there is no such current on the axis. In~either case, as~$P(0,z)=0$, the~integration constant appearing in Equation~(\ref{eq:T-sol}), needs to be zero: $a_2=0$. This leads to the requirement that the quantity appearing inside the bracket in Equation~(\ref{eq:T-sol}) does not become negative. Since the focus of this work is on the logarithmic part of the solution, I set $a_1=0$ and consequently $a_0>0$.

\begin{figure}[H]
    
    \includegraphics[width=0.99\linewidth]{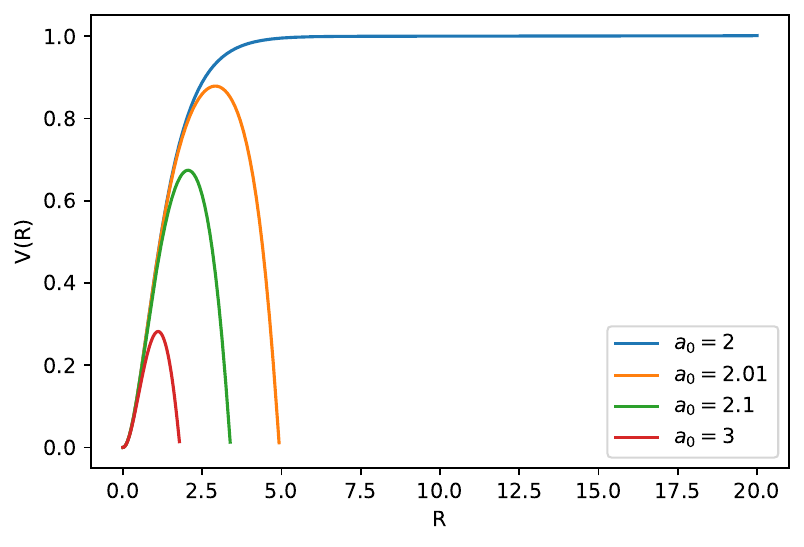}
    \caption{The radial part of the solution $V(R)$, for~$a_0=2$, $a_0=2.01$, $a_0=2.1$ and $a_0=3$. }
    \label{fig:V(R)}
\end{figure}

Since an analytical solution of the ordinary differential equations is far from straightforward, I proceed with numerical integration. The~second term is divided by the radius, thus the integration commences at $R_{in}=10^{-4}$, to~avoid singularities at the origin. Furthermore, I set $V(R_{in}) = V_0$ and $V^{\prime}(R_{in})=V_0^{\prime}$. As~the main focus of the solution is in the logarithmic term, I set $k_1=0$, to~avoid interference with the linear part of the solution. As~explained above $a_0>0$, so that the toroidal field is real-valued. An~interesting feature of this equation is that for $0<a_0<a_{0,crit}$ the solution grows to infinity for large $R$, while for  $a_0>a_{0,crit}$ it reaches a single maximum and then becomes zero again, where due to the logarithmic term the integration cannot proceed any further. The~actual numerical value of $a_{0,crit}$ depends on the choice of the boundary conditions chosen. Here, the~boundary values used are $V_0=10^{-4}$ and $V_0^{\prime}\approx 1.47 \times 10^{-4}$ leading to $a_{0,crit} \approx 2$, for~which $V(R)$ attains its maximum value at unity; therefore, $V^{\prime}=0$ and $\ln V=0$, thus the solution remains constant. This feature of the solution has also been noticed in the plane-parallel geometry; however, in~the cylindrical geometry there is an extra complexity due to the first derivative term, making the value of $a_{0,crit}$ to depend on the choice of the boundary conditions. These behaviors are plotted for three characteristic values of $a_0$ in Figure~\ref{fig:V(R)}. 

Next, by~multiplication of the solutions for $V(R)$ and $Z(z)$, one can obtain the solution for $P(R,z)$. Here, several combinations of solutions are possible, as~for a particular value of $a_0$ qualitatively different behaviors can be found for $V(R)$ and $Z(z)$. 

\begin{figure}[H]
    
    \includegraphics[width=0.84\linewidth]{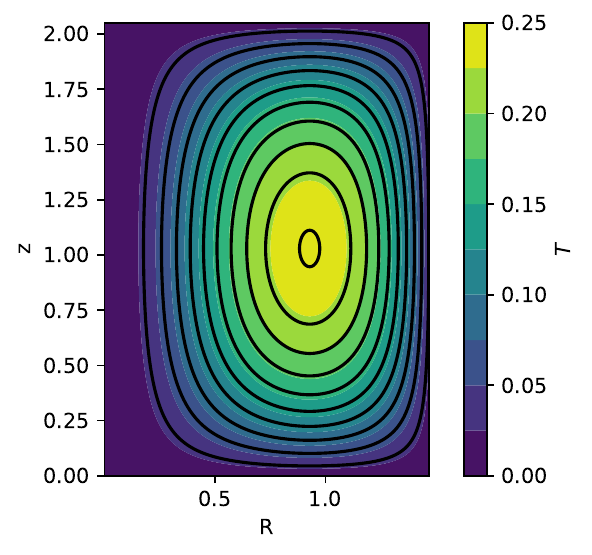}
    \caption{The field line structure for a cylindrical solution with parameters $a_0=3$, $k_1=a_1=0$, and~boundary conditions for the radial part $V_0=V_0^{\prime}=10^{-4}$, while for $Z(0)=0.0$ and $Z^{\prime}(0)=1$. The~black contours correspond to constant values of $P(R,z)$, while the color to constant values of $T(R,z)$. The~solution is bounded both in $R$ and $z$.  }
    \label{fig:a0=3}
\end{figure}

\begin{figure}[H]
    
    \includegraphics[width=0.85\linewidth]{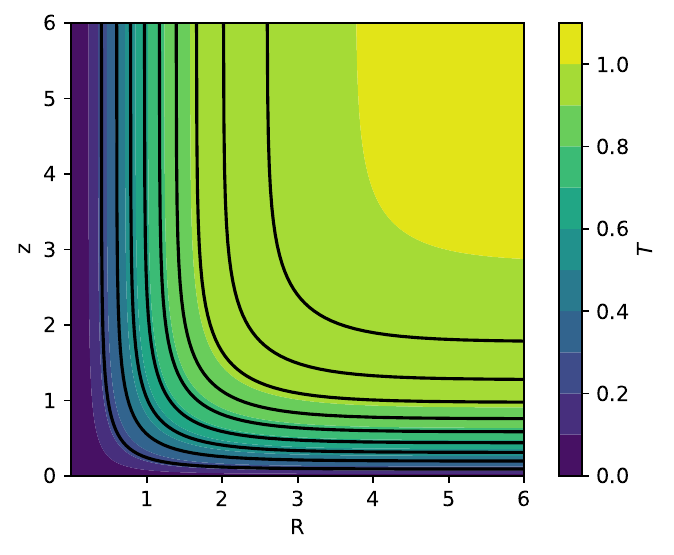}
    \caption{The field line structure for a cylindrical oscillatory solution with parameters $a_0=2$, $k_1=a_1=0$, and~boundary conditions for the radial part $V_0=10^{-4}$, $V_{0}^{\prime} =1.47 \times 10^{-4} $ while for $Z(0)=0$ and $Z^{\prime}(0)=1$. The~black contours correspond to constant values of $P(R,z)$, while the color to constant values of $T(R,z)$. }
    \label{fig:a0=2}
\end{figure}

\begin{figure}[H]
    
    \includegraphics[width=0.99\linewidth]{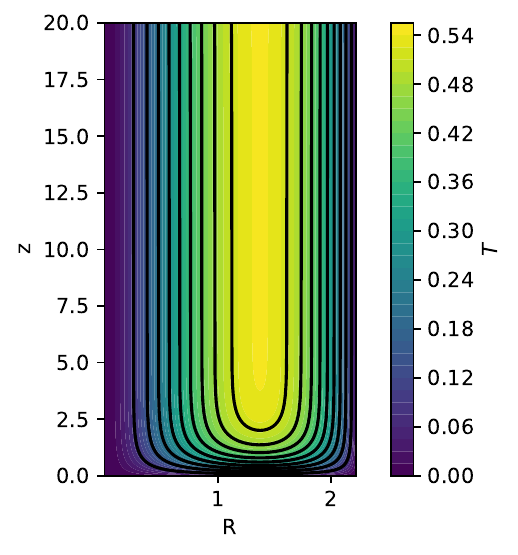}
    \caption{The field line structure for a cylindrical oscillatory solution with parameters $a_0=2$, $k_1=a_1=0$, and~boundary conditions for the radial part $V_0=V_0^{\prime}=10^{-4}$, while for $Z(0)=0$ and $Z^{\prime}(0)=1$. The~black contours correspond to constant values of $P(R,z)$, while the color to constant values of $T(R,z)$. The~solution is bounded in $R$ and becomes constant for large $z$. }
    \label{fig:a0=22}
\end{figure}

In the first example shown ($a_0=3$), the~solution is bounded within a cylinder of given radius and height, see Figure~\ref{fig:a0=3}. Next the field line structure for $a_0=2$ is shown, which attains a constant value both for large  $R$  and $z$, Figure~\ref{fig:a0=2}. Note that here the boundary conditions for $V_0$ and $V_0^{\prime}$ have been deliberately chosen so that the value $a_{crit}=2$ which is also the critical value for the $Z(z)$ function. In~general these two critical values are not expected to be equal. An~example where $V_{0}=10^{-4}$, $V_{0}^{\prime}= 10^{-4}$ and $a_0=2$, has a solution where $Z(z)$ becomes constant, while $V(R)$ becomes the maximum and then goes to zero; see Figure~\ref{fig:a0=22}. 

\unskip
\unskip
\unskip

\section{Discussion}

\label{sec:disc}

Analytical and semi-analytical solutions of force-free magnetic fields are limited due to the complexity and non-linear nature of the systems. A~fully analytical solution in three dimensional systems is the Arnold--Beltrami--Childress, commonly known as ABC fields~\cite{1986JFM...167..353D}, which has been used extensively as initial conditions and benchmark cases in simulations of force-free magnetic fields~\cite{2017JNS....27..721L,2018MNRAS.481.4342N}. In~systems with symmetry, either translational in the plane parallel geometry or azimuthal in the axisymmetric works, there exist the families of linear solutions, for~which fully analytical solutions are feasible. Further analytical solutions, either assuming separation of variables or through constructive methods, have been presented in Cartesian, cylindrical and spherical geometry~\cite{1977ApJ...212..234L,1978SoPh...57...81B,1980SoPh...65..315P,1993ApJ...404..781R}. 

The wide use of numerical methods has allowed the treatment of the highly non-linear differential equations describing force-free fields without the need to appeal to simplifications that lead to analytical solutions. In~this approach, the~form of the source term is prescribed and an appropriate set of boundary conditions is assumed, giving the field or its derivatives at the boundaries of the computational~domain. 

The solutions presented here are based on the assumption of separation of variables. They are not fully analytical, in~the sense that their solutions do not appear in closed form, yet, they simplify to a set of decoupled ordinary equations whose solutions are straightforward using the standard method of 4th order Runge--Kutta~integration. 

The logarithmic family of solutions has some unique properties compared to the known solutions. In~the plane-parallel case it demonstrates four distinct types of~behaviors:
\begin{itemize}
    \item A single maximum solution where the solution starts from a minimum value, typically zero reaches a maximum and then becomes zero again.
    \item An unbounded solution where the solution grows indefinitely.
    \item A solution that reaches an asymptotic value { where the field becomes uniform.}
    \item The periodic solution. In~this case, the~fields are periodic, { while the} magnetic flux function never becomes negative.
\end{itemize}

{ Solutions in cylindrical geometry demonstrate similar features, in~particular, the~cases discussed here have fields exhibiting the following~behaviors:
\begin{itemize}
    \item Fields where the 
 flux function has a single maximum in $z$ and $R$ and become zero at a finite distance, thus, they are confined within a cylinder.
    \item Fields where the flux function becomes constant for large $R$, thus, there is no $z$ component of the field for large $R$. 
    \item Fields where the flux function becomes constant at large $z$, thus, there is no $R$ component there.
    \item Fields where the flux function becomes uniform both at large $R$ and $z$, thus, there is no $R$ and $z$ components at large distances but only $\phi$.
\end{itemize}
}

While some of these behaviors have been noticed in other types of force-free  states, this variety of features makes them  distinct  from the rest. For instance, single maximum and unbounded solutions can be reached through the linear system, either by keeping the oscillatory terms or the exponential ones. Even in the single maximum solution, the~solution needs to be confined through a current sheet, as~while the flux function becomes zero at some point, its derivative does not, thus there is a magnetic field discontinuity there. On~the contrary, the~logarithmic solution may reach an asymptotic constant value in the $z$ direction either in planar or axial symmetry and even possibly in the $R$ direction. It is remarkable that solutions of magnetic fields that are not confined by current sheets require an additional thermal pressure \cite{2010MNRAS.409.1660G,2012MNRAS.419.3048G}, whereas in this configuration this is feasible through the logarithmic term. Finally, the~solution that exhibits periodic behavior has a $B_z$ magnetic field that is always pointing in the same direction. Comparing them to the periodic fields that correspond to linear solutions, they never become negative and the flux function always remains positive and so does the field that points in the symmetry direction. A~clear distinction between the logarithmic solution and other classes of solutions is the fact that in the logarithmic solution the flux function does not become negative. A~possible extension could be found if the expressions depend on the absolute values of the flux functions, but~this is left for future work, as~it may be possible that some discontinuities in the derivatives may~appear. 

Given the freedom on the $\alpha$ parameter that prescribes the ratio of magnitude of the current to the magnetic field, one can choose a functional form for $\alpha$ which is constant along the magnetic field lines and construct a solution. However, even in this case, it is not guaranteed whether this solution will be the one that will be adopted by nature. This question still remains unresolved, even though some physically motivated arguments have been proposed. In~the case of a force-free magnetic field of given helicity in a closed system, the~minimum energy configuration will be the one that has a constant $\alpha$ \cite{1958PNAS...44..489W}. However, several systems of astrophysical interest are not force-free { in the entire domain}, as~the magnetic field emerges from a non force-free region, i.e.,~the surface of a star. Similarly, in~configurations where critical surfaces are included, such as the pulsar magnetosphere that contains the light-cylinder, it is possible to determine the expression of $\alpha$ that corresponds to the force-free configuration for which the magnetic field smoothly crosses the light-cylinder, in~which case, the~actual solution is clearly non-linear~\cite{1999ApJ...511..351C}. Similarly, in~models of the solar magnetic field non-linear constructive solutions provide successful results~\cite{2018SSRv..214...99Y}. Thus, the~family of solutions in hand, where the source is related to the logarithm, widens the horizons of solutions that are tractable with methods not as complicated as the full numerical solution of the non-linear force-free~equations. 

While these solutions require a special dependence of the current on the magnetic flux, some of their properties are potentially useful to astrophysical systems. In~the cylindrical geometry, the~solutions where the $R$ component of the magnetic field vanishes at large $z$, such as shown in Figure~\ref{fig:a0=22}, are reminiscent of the magnetic tower configurations~\cite{2003MNRAS.341.1360L,2006MNRAS.369.1167L} that have been used extensively in jet models~\cite{2004MNRAS.351L..89G,2017ApJ...839...14G} and also in laboratory plasma experiments~\cite{2013NJPh...15l5008S} and simulations~\cite{2007Ap&SS.307...17C}. Note that the magnetic tower model is based on energy arguments, rather than an exact solution, whereas the logarithmic solution provides a suitable exact force-free state, if~the origin of the jet lies at $z\gg 1$ and the top of the jet at $z=0$, for~the solution presented here. An~other type of solution of possible astrophysical interest is the one shown in Figure~\ref{fig:a0=2} that could envision a jet due to a magnetic field collimated on the axis, where field emerges from an extended disk that lies on the horizontal plane. This solution is confined by a toroidal field $B_{\phi}$ that is force-free and at large distances drops as $R^{-1}$. 

\section{Conclusions}
\label{sec:Con}

This work has introduced a new class of solutions for force-free magnetic fields in Cartesian geometry where plane-parallel symmetry has been imposed and in cylindrical geometry where axial symmetry has been used. This solution allows the separation of variables in either case, due to the elementary property of logarithms that the logarithm of the product of two functions is the sum of their logarithms. The~fact that this highly non-linear equation separates to a system of still non-linear, yet decoupled, ordinary differential equations, allows the thorough exploration of their properties via changing the boundary conditions and the proportionality parameters appearing. These solutions demonstrate some quite remarkable properties, as~they transition smoothly to uniform magnetic fields in the plane parallel case, and~fields corresponding to constant poloidal current~function. 

Despite their interesting properties, it is not self-evident whether such solutions would correspond to a natural and  spontaneous force-free { state} reached by a physical system. However, they can be useful initial conditions for studies of systems where a force-free has a non-trivial geometry, but~is confined by a uniform field at large distances. Because~of that, they can be used as a useful set of initial conditions and benchmark cases for numerical simulations. { Moreover, some of their properties are relevant to astrophysical systems and resemble configurations that have been used to model astrophysical jets in the form of magnetic towers and fields confined round an axis.} Finally, it has been shown that the logarithmic term can co-exist with a linear term in the electric current distribution. While some solutions of mixed type: logarithmic and linear, have been presented in the current paper, these fields clearly merit a detailed study focusing on such combinations and is reserved for future~work.


\vspace{6pt} 




\funding{KNG acknowledges funding from grant FK 81641 ``Theoretical and Computational Astrophysics'', ELKE. This work was supported by computational time granted by the National Infrastructures for Research and Technology S.A. (GRNET S.A.) in the National HPC facility---ARIS---under project ID pr015026/simnstar.}

\dataavailability{Numerical data related to this study are available upon request.} 


\conflictsofinterest{The author declares no conflicts of interest.} 

\appendixtitles{yes} 
\appendixstart
\appendix
\section[\appendixname~\thesection]{The Form of $\alpha$}\label{appA}


In the approach followed in the paper, Equation~(\ref{eq:FF1}) was used to derive the final equation in plane-parallel and cylindrical geometry. The~same results can be reached if Equation~(\ref{eq:CurlB-B}) is~used.

In the plane-parallel geometry, taking the curl of the magnetic field and using elementary vector calculus identities, one obtains 
: 
\begin{eqnarray}
    \nabla \times \left(\nabla \Psi \times \nabla z +B_{z}\nabla z\right) = -\nabla^2 \Psi \nabla z +\nabla B_z \times \nabla z\,,
\end{eqnarray}
then by virtue of Equation~(\ref{eq:CurlB-B}), one finds:
\begin{eqnarray}
    -\left(\nabla^2 \Psi +\alpha B_z\right)\nabla z +\left( \nabla B_z -\alpha \nabla \Psi\right) \times \nabla z={\bf 0}\,.
\end{eqnarray}

A comparison of the first bracket of the above expression with Equation~(\ref{eq:PDE}) leads to
\begin{eqnarray}
    \alpha = B_{z}^{\prime}\,.
\end{eqnarray}
inally, using expression \eqref{eq:Bz}, the~value of $\alpha$ is written in terms of $\Psi$ and the various parameters in the following form:
\begin{eqnarray}
    \alpha =\frac{2\Psi\left(-c_0 \ln \Psi +c_1\right)}{\left[2\Psi^2 \left(-c_0 \ln \Psi +\frac{c_0}{2} +c_1\right)+c_2\right]^{1/2}}\,.
\end{eqnarray}

A similar procedure can be followed for the axially symmetric geometry. The~force-free equation takes the form:
\begin{eqnarray}
    R\frac{\partial }{\partial R}\left(\frac{1}{R}\frac{\partial P}{\partial R}\right) +\frac{\partial^2 P}{\partial z^2} = -\alpha T\,.
\end{eqnarray}

Therefore, it is:
\begin{eqnarray}
    \alpha = T^{\prime}\,.
\end{eqnarray}

Given the expression for $T^2$ from Equation~(\ref{eq:T-sol}), the~form for $a$ in terms of $P$ and the various parameters is:
\begin{eqnarray}
\alpha =\frac{2P\left(-a_0 \ln P +a_1\right)}{\left[2P^2 \left(-a_0 \ln P +\frac{a_0}{2} +a_1\right)+a_2\right]^{1/2}}\,.
\end{eqnarray}


\begin{adjustwidth}{-\extralength}{0cm}
\reftitle{References}



\PublishersNote{}

\end{adjustwidth}

%


\end{document}